# Interfacial amplification for graphene based position-sensitive-detectors

Wen-Hui Wang[1], Ru-Xia Du[2], Xi-Tao Guo[1], Jie Jiang[1], Wei-Wei Zhao[3], Zhong-Hua Ni[3], Xin-Ran Wang[4], Yu-Meng You[5] and Zhen-Hua Ni[1]

[1]State Key Laboratory of Bioelectronics, Department of Physics, Southeast University, Nanjing 211189, China.

[2]Department of Basic Teaching, Nanjing Tech University Pujiang Institute, Nanjing 211134, China.

[3]Jiangsu Key Laboratory for Design and Fabrication of Micro-Nano Biomedical Instruments, School of mechanical engineering, Southeast University, Nanjing 211189, China.

[4]National Laboratory of Solid State Microstructures, School of Electronic Science and Engineering, and Collaborative Innovation Center of Advanced Microstructures, Nanjing University, Nanjing 210093, China.

[5]Ordered Matter Science Research Center, Southeast University, Nanjing 211189, China.

Correspondence: Zhen-Hua Ni, Email: zhni@seu.edu.cn

**Position-sensitive-detectors (PSDs) based on lateral photoeffect has been widely used in diverse applications[1-9], including optical engineering, aerospace and military fields. With increasing demand in long distance, low energy consumption, and weak signal sensing systems, the poor responsivity of conventional PSDs has become a bottleneck limiting its applications, e.g. silicon p-n or p-i-n junctions[2-5], or other materials and architectures[6-10]. Herein, we present a high performance graphene based PSDs with revolutionary interfacial amplification mechanism. Signal amplification in the order of ~$10^4$ has been demonstrated by utilizing the ultrahigh mobility of graphene and long lifetime of photo-induced carriers at the interface of $SiO_2$/Si. This would improve the detection limit of Si-based PSDs from μW to nW level, without sacrificing the spatial resolution and response speed. Such interfacial amplification mechanism is compatible with current Si technology and can be easily extended to other sensing systems [11,12].**

Figure 1a shows the representative schematic and the working principle of the graphene based PSD, which contains a two-terminal graphene transistor deposited on the top of 300 nm thick $SiO_2$/lightly p-doped (1-10 Ω cm) Si substrate. The localized interface states[13] such as positive charge states with energies within the silicon band gap exist at the oxide-silicon interface, induce a negative depletion layer (-) in the silicon near the interface, making the surface energy bands bend downwards. This will lead to the formation of the intrinsic built-in electric field[13,14] with the direction from the interface to bulk Si, as shown in Figure 1b. Under the illumination of a point light source, as shown in Figure 1a, electron-hole pairs will be generated inside or within the minority-carrier diffusion length of the depletion region[6,7], and then be separated by the built-in electric field. In such case, the photo-induced electrons would accumulate at $SiO_2$/Si interface and diffuse laterally. When reach equilibrium, electrons will unevenly but steadily distribute at the interface, following a conventional lateral photoeffect-like behavior[3-9]. The electrons diffuse to the region under graphene will lead to an effective gate effect, consequent changes the hole concentration and channel current through capacitive coupling. This would result in an ultrahigh gain or amplification, namely interfacial amplification[14], which originates from the recirculation[15] of holes in the graphene channel during the lifetime of electrons that accumulate at $SiO_2$/Si interface. The interfacial amplification in our system is analog to photoconductive gain, which is determined by the ratio of lifetime and transit time of carriers between the two electrodes[15]. Here, the gain[16] or amplification is defined as $G = \tau_l / \tau_t$, here, $\tau_l$ is the lifetime of photo-induced carrier at the $SiO_2$/Si interface, $\tau_t$ is the transit time in the graphene channel. The extreme short transit time $\tau_t$ due to ultrahigh mobility of graphene [17,18] and long lifetime of carriers $\tau_l$ at $SiO_2$/Si interface would result in high gain G in our device. Such an interfacial amplification process behaves like a built-in amplifier, but it would not increase the noise level of the system (the signal noise in the graphene channel), so that the sensitivity of the device would be dramatically enhanced. The quantity of the accumulated electrons under the graphene channel varies with the incident light position, suggesting that different photocurrent will take place, and also the capability of ultrasensitive position detections of the light point by considering the high gain of interfacial amplification mechanism.

The position sensitive photoresponse characteristics of the device were recorded with laser focused on the device (514 nm, spot size ~1 μm) under a fixed DC bias, $V_{DC}$ = 1 V. Figure 1c shows a typical photoresponse for our device, where the photo-switching characteristics of the graphene photodetector are plotted with light (395 nW) incident on the substrate with different distances away from graphene channel. It is clear that considerable photocurrent generated and its magnitude declines significantly with increasing distance from the laser spot to graphene, as also shown in Figure 1d. According to the non-equilibrium carrier diffusion theory, the amount of carriers decrease exponentially with the distance away from the laser spot due to recombination in the transport process, so does the photocurrent in the graphene channel. The exponential fitting curve in Figure 1d tally with the experimental photocurrent well, confirmed the mechanism of lateral photoeffect discussed above.

Figure 2a shows the position dependent photocurrent response of the PSD under different light power down to nW. We note that, even at extremely low incident power of 1.4 nW, a significant steady decrease of the photocurrent with the increasing distance is observed, suggesting excellent position sensitivity in our device to weak light signal. Furthermore, as seen in Figure 2b, a critical parameter in this system, the responsivity can be derived as high as ~$10^3$ A $W^{-1}$. In conventional PSDs with silicon (Si) p-n or p-i-n junctions[2-5], photoexcited electron-hole (e-h) pairs are separated by built-in electric field existed at the junction, giving rise to a lateral potential gradient[3] between the illuminated and non-illuminated zones, thus the photo-induced carriers will diffuse and directly collected by two electrodes at both ends of surface layer. These devices are appealing for good linearity[6-9] and fast response, but with relatively poor photoresponsivity (< 1 A/W). This is mainly because no effective gain mechanism exists in these structures. In our device, the interfacial amplification would result in a huge responsivity and improve the detection limit of PSDs from mW/μW of conventional Si PSDs[3-9] to nW, demonstrating the potential of this device for weak signal detection. The laser spot size dependence of the photocurrent was also measured at 500 μm away from graphene under different illumination intensity, as shown in Supplementary Fig. S3. The photocurrent keeps constant with variation of the spot size, in good accordance with the characteristic of "independent of the incident light shape" of PSD. It should also note that the PSD can work in broad-band with wavelength range from 320 nm to 1100 nm as silicon is the photosensitive material.

Next, the time response of graphene based PSD was studied by using an acoustic optical modulator with frequency of 10 kHz to switch the light (~400 nW) on or off at 5 μm away from the graphene channel. As shown in Figure 2c, the rise ($\tau_{on}$) and fall ($\tau_{off}$) time are ~400 ns and ~1.2 μs, respectively, where the rising and falling parts of the curves are fitted using a single exponential function. Since the recombination processes are statistically random, the photo-induced carrier lifetime $\tau_l$ at the interface is close to $\tau_{off}$ ($10^{-6}$ s), while the transit time $\tau_t$ in the graphene channel is in the order of $10^{-10}$ s at 1 V bias for our device (the mobility is ~8000 $cm^2V^{-1}s^{-1}$ as shown in Supplementary Fig. S1(b)), leading to the amplification or gain ($G = \tau_l/\tau_t$) of ~$10^4$. We have carried out control experiments by substituting high mobility graphene with low mobility $MoS_2$[19] (Supplementary Figure S4), or substituting lightly doped $Si/SiO_2$ substrate with heavily doped $Si/SiO_2$ substrate (Supplementary Figure S5), the photoresponse as well as gain is negligible. These results suggest that the ultrahigh mobility of graphene and the long lifetime of electrons at the interface of lightly doped $Si/SiO_2$ substrate are the two key factors for interfacial amplification.

The high speed response of the device could be attributed to the fast separation and diffusion of the photo-induced carriers at the lightly doped $Si/SiO_2$ interface. For better understanding of this kinetics, we analyzed the position dependent rise time with different illumination intensities, as shown in Figure 2d. The tested $\tau_{on}$ is almost flat and fixed on ~500 ns within 20 μm under 3.4 μW, 100 μm under 30 μW, indicating that the diffuse time of carriers in $Si/SiO_2$ interface or bulk Si is short and can be negligible within the corresponding distance ranges. However, when the distance excesses the threshold value, the diffusion time of carriers at the $SiO_2/Si$ interface can no longer be ignored and dramatically increase with distance. This can be explained by the change of concentration gradient of photo-induced carriers. Under low light intensity or the position far away from incident point, few photo-induced carriers exist at the interface and induce a small lateral potential gradient, which will increase diffusion time of the carriers. Despite the increase, the response speed of our PSD is still comparable with conventional Si-based PSDs, suggests that the interfacial amplification could greatly enhance the sensitivity of PSD without sacrificing the high response speed.

The kinetics of photo-induced carriers at the lightly p-doped $Si/SiO_2$ substrate was further characterized by using capacitance–voltage (C–V) measurements. For metal-insulator-semiconductor (MIS) structure, the capacitance equals to the capacitance of oxide layer and silicon surface space charge layer in series: $1/C = 1/C_0 + 1/C_s$, Where $C$, $C_0$ and $C_s$ are the measured oxide layer and silicon surface space charge layer capacitance, respectively. Figure 3a displays the C-V characteristic curves in dark and different illumination intensity at 5 kHz. It can be seen that the flatband voltage[20] (VFB) is ~ -4 V, as the normalized flatband capacitance is ~0.95 for our substrate. This signifies that, under no bias, the energy band bend downwards at Si surface and there is a built-in electric field, in good agreement with the analysis shown in Figure 1b. The negative shift and increase of the capacitance minimum value under illumination suggest the injection of electrons at the interface of $SiO_2/Si$ with the increase of light power (more details about the C-V measurements can be seen in Supplementary Fig.

S6). Figure 3b presents the relationship between the photo-induced specific capacitance (ΔC) and the illumination intensity at V = 0 V. We found that the photo-induced specific capacitance grows and finally saturates with the increase of incident power. It is because that the accumulated electrons progressively counteract the built-in electric field until completely balanced. The capacitance therefore reaches a steady state, which is in consistence with the saturation of photocurrent with the increase of incident power as mentioned above. In addition, we have also measured the capacitance of heavily doped silicon substrate, no response was found under the illumination. Hence, it is demonstrated that the lightly doped Si/SiO$_2$ interface plays a central role in the photosensitive behavior of our device.

To prove the capability of position detection of graphene based PSD, a one-dimensional PSD was prepared by using two graphene devices (S1, S2) on the same substrate separated by 10 μm. Figure 4a shows position dependence of photoresponse of the two devices under 50 nW incident light, and the optical image of the PSD are displayed in the inset of Figure 4a. The position sensitivity of the PSD can be observed from the increased or decreased photocurrent as incident light go toward or away from the graphene, and the position of incident could be deduced by the photoresponse ratio $(I_2-I_1)/(I_1+I_2)$ of the two devices, as shown in Figure 4b. It can be clearly seen that the ratio is zero at position near middle of the PSD, and performs well linear in detecting range. Considering the detection limit of photocurrent (~0.5 μA), the spatial position resolution should be less than 1 μm. Besides, two-dimensional spatial mapping of photocurrent under the illumination of 400 nW laser is also demonstrated in Figure 4c. The overlapping distance dependence of photoresponse at different directions, as shown in Figure 4d, verified that diffusion of the photo-induced carriers is isotropic. The uniform and protected SiO$_2$/Si interface guarantees almost the same diffusion, which confirmed the potential utilization for ultra-fast weak signal two-dimensional PSDs.

The graphene based PSDs reported in this work shows high responsivity and fast response time, which provides important advances and new development opportunity for future weak signal sensing.

**MATERIALS AND METHODS**

The single layer graphene sample is mechanically exfoliated from highly oriented pyrolytic graphite (HOGP) onto the top of 300 nm thick silicon oxide/lightly p-doped (1-10 Ω cm) silicon (SiO$_2$/Si) substrate. The source and drain electrodes (Ni (5 nm)/Au (50 nm)) were patterned by electron-beam lithography (FEI, FP2031/12 INSPECT F50), thermal evaporation (TPRE-Z20-IV), and lift-off processes. The optical image of the device as well as the Raman spectrum suggests the monolayer thickness of the graphene sample (see Supplementary Fig. S1(a)). The transfer curve of the graphene device is also obtained and proves its high quality (Fig. S1(b)). Photoresponse characteristics of the devices are measured using a Keithley 2612 analyzer. All the measurements were performed in air at room temperature. An Ar+ laser with wavelength of ~514 nm is employed to attain the photocurrent response, which is focused on the sample with a 50 x objective (NA= 0.5) and the spot size of light is ~1 um. In the response time measurement, light was modulated with an acoustic optical modulator (R21080-1DS) at frequency of 10 kHz. A digital storage oscilloscope (Tektronix TDS 1012, 100 MHz/1 GS/s) is used to measure the transient response of photocurrent. The photoresponse spatial scanning/mapping is obtained by moving the two dimensional stage and obtaining the photocurrent point by point. The C–V characteristics of the substrate were measured by using Keithley 4200 semiconductor characterization system.

## AUTHORS' CONTRIBUTIONS

Zhen-Hua Ni conceived the project. W H W carried out device fabrication, electrical and photoresponse measurements and data analysis. X T G. and J J performed control device fabrications. W W Z and Zhong-Hua Ni helped on the C-V measurements. Zhen-Hua Ni, W H W, R X D, X R W and Y M Y co-wrote the paper with all authors contributing to the discussion and preparation of the manuscript.

## ACKNOWLEDGEMENTS

This work was supported by the National Key Research and Development Program of China (No. 2017YFA0205700), NSFC (61422503 and 61376104), the Fundamental Research Funds for the Central Universities, and Research and Innovation Project for College Graduates of Jiangsu Province No. KYLX15_0111.

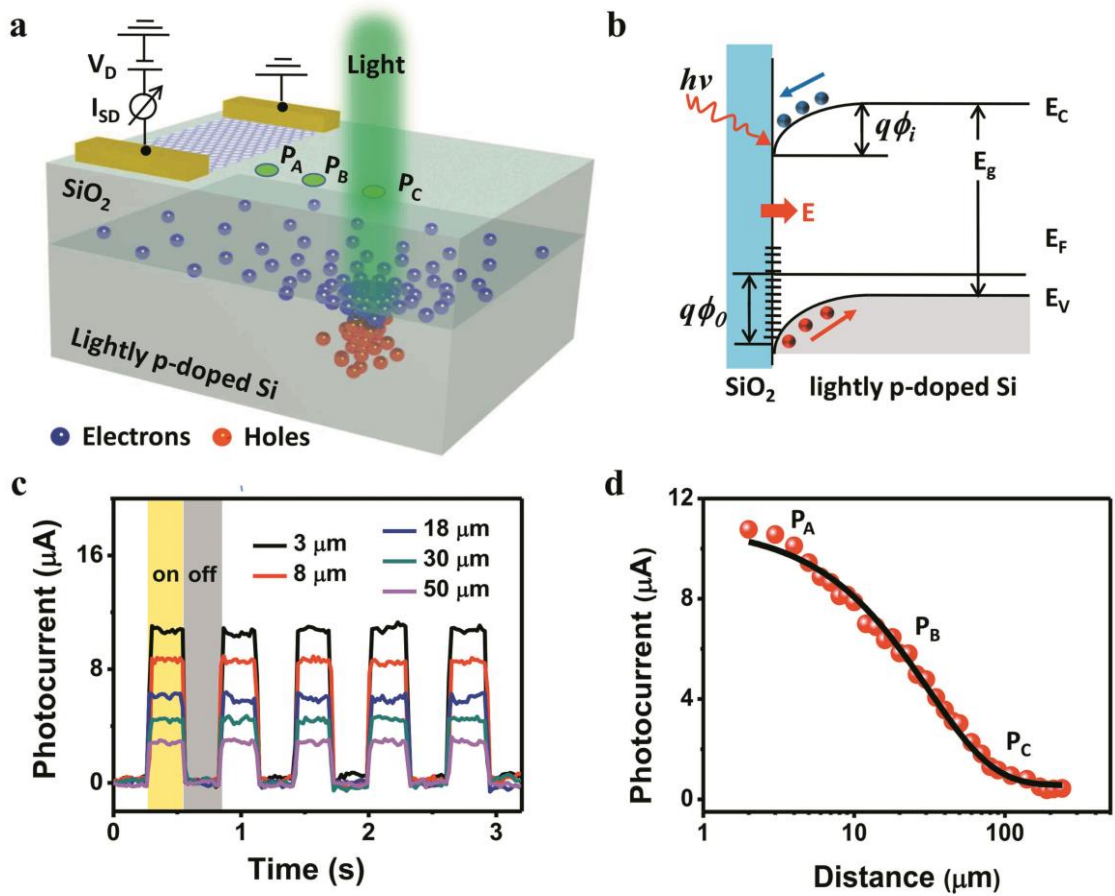

**Figure 1** Schematic of graphene based PSD with interfacial amplification. (**a**) Schematic diagram of the graphene-based PSD. Incident light creates electron–hole pairs in the lightly p-doped silicon. Holes remain in the silicon, while Electrons are accumulated at $SiO_2$/Si interface and diffuse laterally. (**b**) Schematic of energy band diagrams of the lightly p-doped silicon/$SiO_2$ interface with positive localized states. (**c**) Photo-switching characteristics of the graphene photodetector at different distance (from the laser spot to graphene device) using 395 nW incident light. (**d**) The spatial dependence of the photocurrent. The black solid line is the exponential fitting curve of the experimental data.

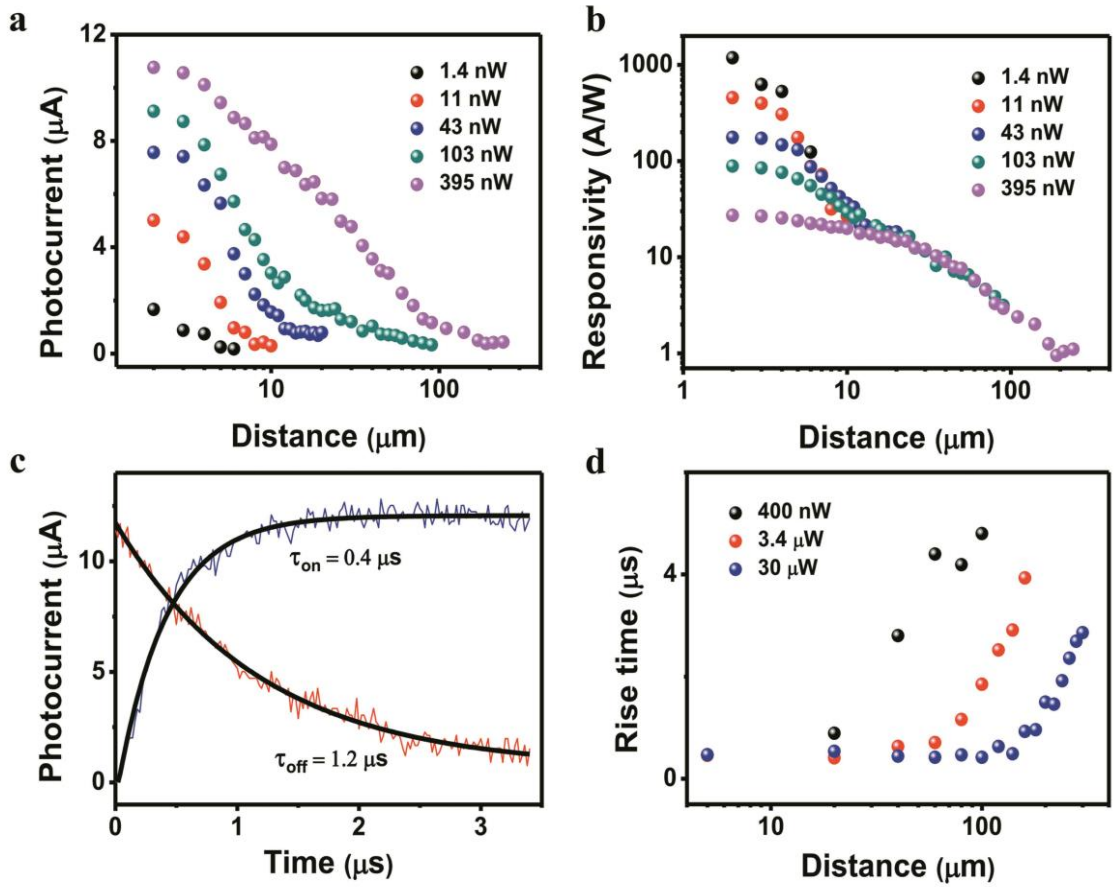

**Figure 2** Performance of the graphene based PSD. (**a, b**) Photocurrent and responsivity as a function of distance at $V_D = 1$ V under different weak light powers using 514 nm wavelength. (**c**) The transient response of the device at 5 μm away from graphene, and light switched on or off by an acoustic optical modulator with frequency of 10 kHz. Light power = ~400 nW, $V_D = 1$ V. (**d**) The distance dependence of the rise time of the device under different illumination intensity.

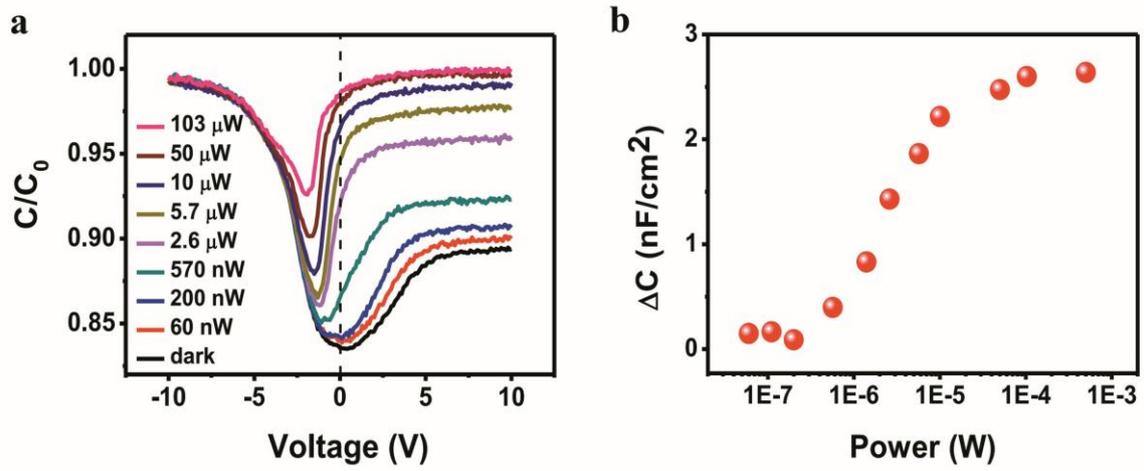

**Figure 3** The capacitance characteristics of lightly p-doped SiO$_2$/Si. (**a**) C-V characteristic curves in dark and different illumination intensity at 5 kHz. (**b**) Power dependence of the photo-induced specific capacitance at V = 0 V.

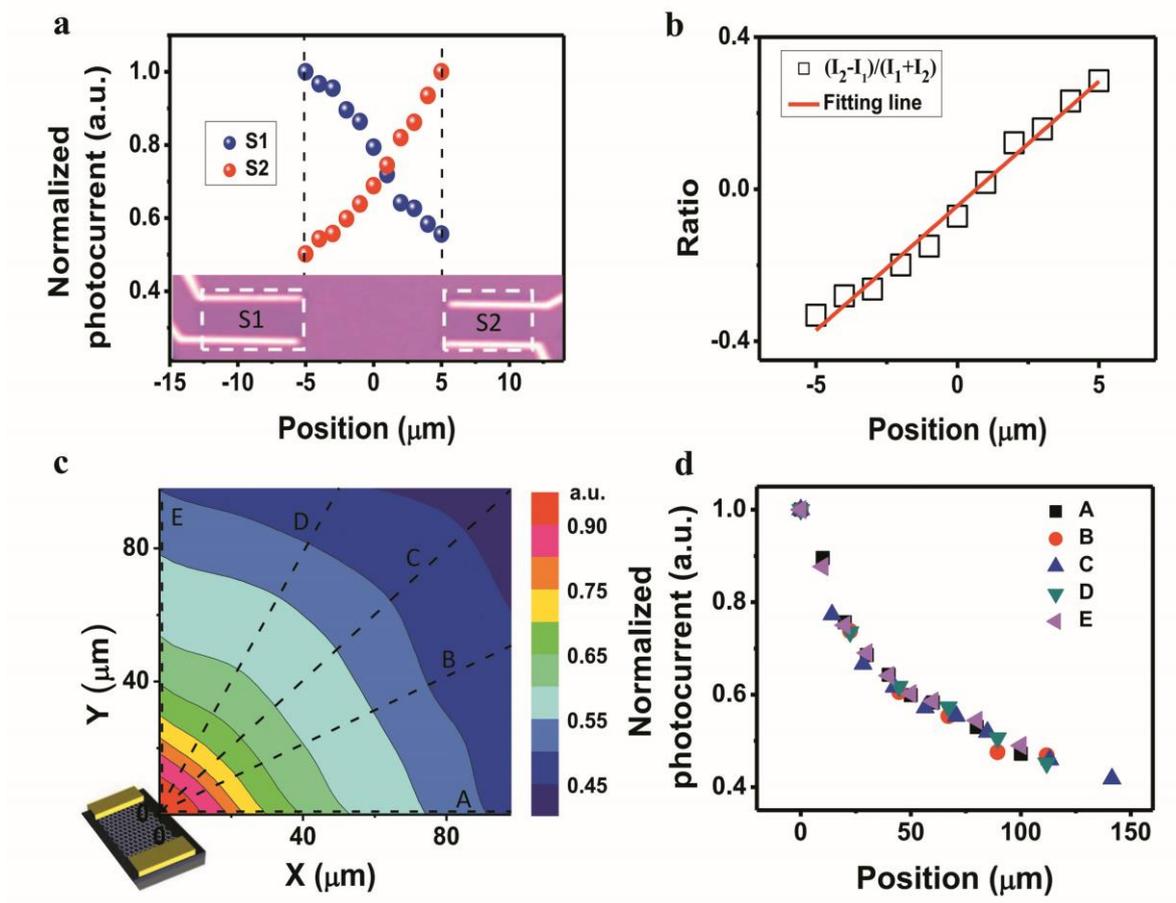

**Figure 4** The representative models of the PSDs. (**a**) Position dependence of photoresponse of one-dimensional PSD prepared by using two graphene devices on the same substrate under 50 nW incident light. The distance between two devices is 10 μm. Optical image of the PSD is displayed in the inset. (**b**) Photocurrent ratio $(I_2-I_1)/(I_1+I_2)$ of two devices as function of the position. (**c**) Two-dimensional spatial mapping of photocurrent of the graphene-based photodetector under 400 nW incident light. (**d**) Photocurrent line scanning as labeled by directions A, B, C, D, E in (**c**).

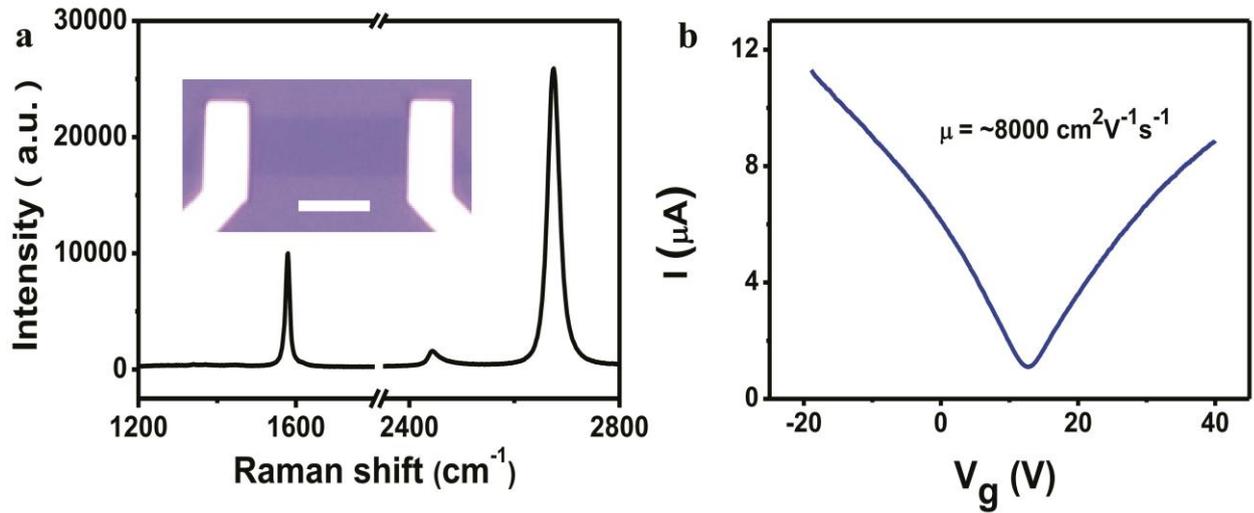

**Supplementary Figure S1** | (a) Raman spectrum and optical image of the graphene device on lightly p-doped Si/SiO$_2$ substrate. The scale bar in optical image is 5 μm. (b) Gate dependence of the drain current for a typical graphene transistor. The monolayer graphene is characterized by optical contrast and Raman spectroscopy [1]. The G band and 2D band (with a full width at half maximum of 27.7 cm$^{-1}$) locate at ~1580 and ~2675 cm$^{-1}$, respectively, and the ratio of I$_{2D}$/I$_G$ is ~2.5 [1]. The field-effect mobility (μ) is calculated by the formula: $\mu = \dfrac{g_m L}{C_g V_{DS} W}$, where $g_m$ is transconductance of graphene (can be obtained from the transfer curve), $L$, $W$ is length and width of graphene channel, $V_{DS}$ is the bias voltage used in transfer curve measurement, $C_g$ is specific capacitance of 300 nm SiO$_2$.

[1] Z. H. Ni, H. M. Wang, J. Kasim, H. M. Fan, T. Yu, Y. H. Wu, Y. P. Feng, and Z. X. Shen, "Graphene Thickness Determination Using Reflection and Contrast Spectroscopy," Nano Lett. 7, 2758–2763 (2007).

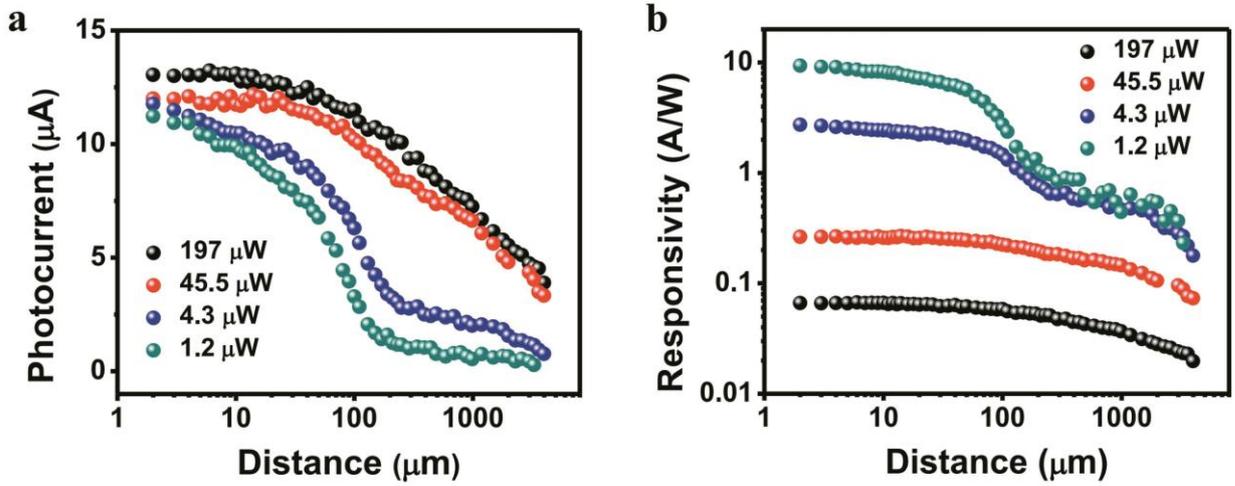

**Supplementary Figure S2** | **(a)** Photocurrent and **(b)** photoresponsivity at $V_D = 1$ V of the device as a function of position under different strong light power. The wavelength is 514 nm. We note that photocurrent will reach saturation and the photoresponsivity drop under excessive high intensity. This saturation behavior could be understood as the reversed electric field induced by the accumulated electrons at the $Si/SiO_2$ interface completely balance to the equilibrium built-in field. The increase of electron-hole recombination rate under high power illumination could be another reason.

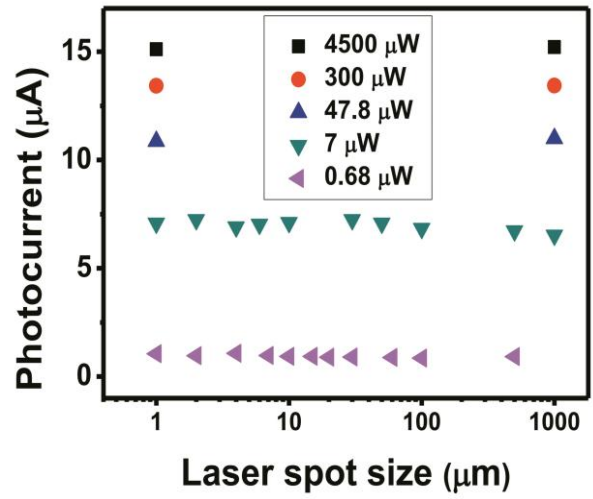

**Supplementary Figure S3 |** Laser spot size dependence of the photocurrent at 500 μm away from graphene under different incident power.

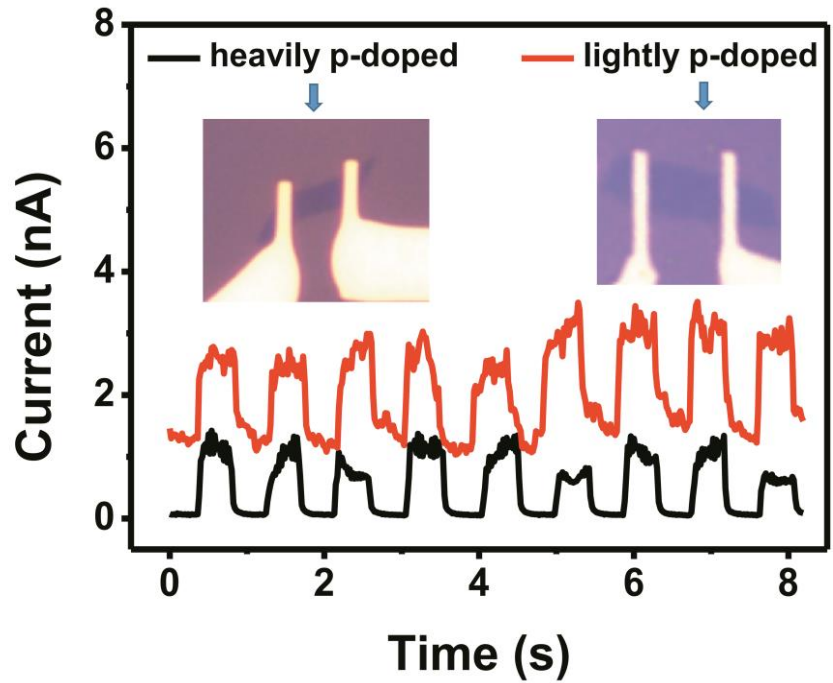

**Supplementary Figure S4** | The photoresponse of MoS$_2$ devices under light illumination ($\lambda$=514 nm, P =1.5 μW, V$_D$=1 V, V$_G$=0 V), the insets show the optical images of monolayer MoS$_2$ devices on lightly and heavily p-doped Si/SiO$_2$ substrates, respectively. We have shown that by substituting high mobility graphene with MoS$_2$ (another two dimensional materials but with relatively low mobility of ~0.1-10 cm$^2$/Vs), the photoresponse is only ~1 nA for 1.5 uW light illumination, corresponding to responsivity of ~0.7 mA/W. The light position sensitivity in this case could be ignored. This demonstrates the effect of high mobility of graphene on the high gain of the system.

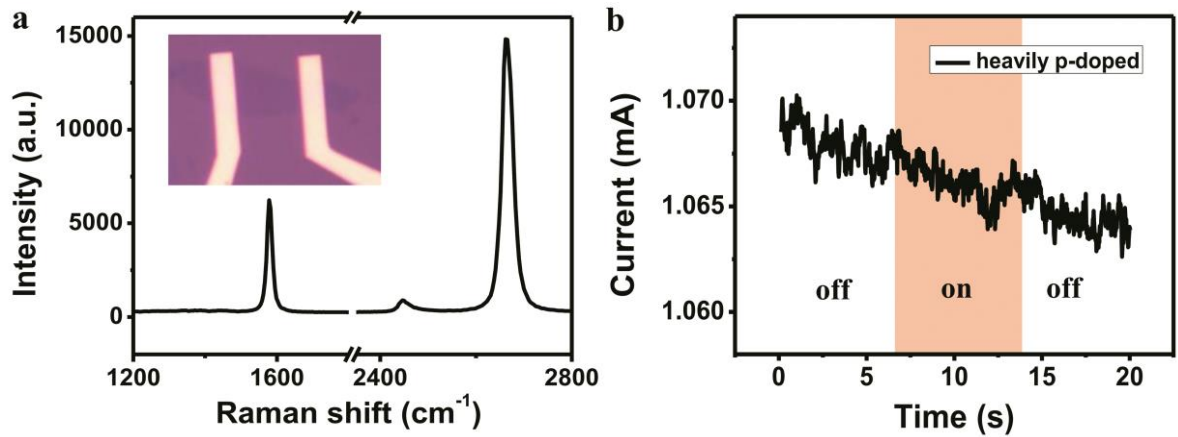

**Supplementary Figure S5 |** (a) Raman spectrum and optical image of the graphene device on heavily p-doped Si (~$10^{-3}$ Ωcm)/SiO$_2$ substrate. (b) No obvious photocurrent could be observed within the measurement resolution of the electronics when switching on and off the light ($\lambda$= 514 nm, P =1.5 μW, $V_D$=1 V). This is because the interfacial amplification in such case is negligible due to the very short lifetime of the photogenerated carriers in heavily doped silicon [2].

[2] A. Cuevas and D. Macdonald, "Measuring and Interpreting the Lifetime of Silicon Wafers," Sol. Energ. 76, 255–262 (2004).

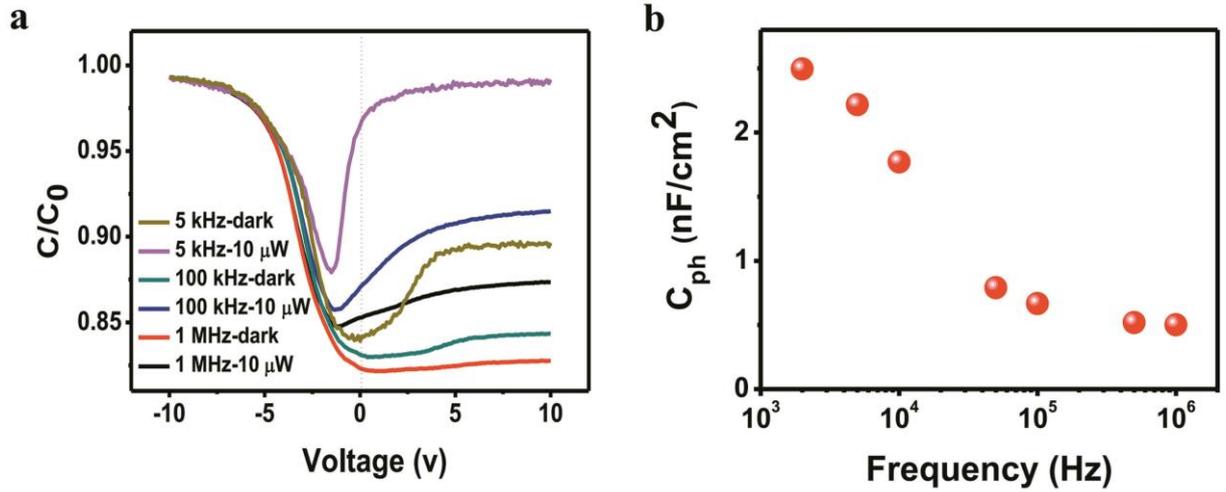

**Supplementary Figure S6 | The capacitance characteristics of the experimentally tested substrate (lightly p-doped SiO$_2$/Si). (a)** C-V characteristic curves in dark and illumination at different frequency from 5 kHz to 1 MHz. **(b)** The frequency dependence of the photo-induced specific capacitance at V = 0 V.

**C-V characteristics**

For metal-insulator-semiconductor (MIS) structure, the capacitance equals to the capacitance of oxide layer and silicon surface space charge layer in series: $1/C = 1/C_0 + 1/C_s$, where $C$, $C_0$ and $C_s$ are the measured oxide layer and silicon surface space charge layer capacitance, respectively. Supporting Information Figure S4a displays the C-V characteristic curves in dark and illumination at different frequency from 5 kHz to 1 MHz. In the C-V curve at 5 kHz in dark, we note that, the capacitance is in an upward slope at V > 0 V, indicating that the silicon surface is in inversion layer. Moreover, the generation and recombination of the electrons in inversion layer cannot follow the ac signal at high frequency, as a result, in this case, the capacitance is mainly determined by the charge change in depletion layer. Additional information of the substrate could be also obtained from the C-V curve, e.g., the maximum depletion layer thickness. The depletion layer thickness will reach the maximum value at strong inversion state, corresponding to the region of V > 5 V. According to the formula,

$$\frac{C_{min}}{C_0} = \frac{1}{1 + \frac{\varepsilon_{r0} x_{dm}}{\varepsilon_{rs} d_0}}$$

Where $C_{min}$ is minimum capacitance, $x_{dm}$ is maximum depletion layer thickness, $d_0$ is thickness of silicon oxide, $\varepsilon_{rs}$ and $\varepsilon_{r0}$ are relative dielectric constant of silicon and silicon oxide, respectively. The value of $\frac{C_{min}}{C_0}$ is 0.826 for our substrate, therefore, the maximum depletion layer thickness is calculated to be ~200 nm, whereas the inversion layer thickness generally in range 1-10 nm. The maximum depletion layer thickness is less than the penetration depth in the substrate for 514 nm laser. This suggests that the photo-induced carriers are generated in both depletion layer and bulk silicon. In addition, C-V characteristic could be used to effectively monitor the carrier concentration in substrate. Comparing the C-V curve in dark and illumination, as seen in this figure, we noticed that the C-V characteristic transits from high frequency to low frequency under illumination. This is due to the increase of recombination and generation rate of carriers. The frequency dependence of the photo-induced specific capacitance was extracted at V = 0 V (Supporting Information Figure S4b), and the capacitance increases obviously with the decrease of the frequency, which is due to that the carriers cannot follow the ac signal at high frequency. Thus, relatively low frequency of 5 kHz was adopted in main text.